\documentclass[12pt]{article}
\pdfoutput=1

\usepackage{putex}

\usepackage{mathtools}
\usepackage{mathrsfs}
\usepackage{graphicx}
\usepackage{caption}
\usepackage{subcaption}
\usepackage{epstopdf}
\usepackage{enumerate}
\usepackage{cite}
\usepackage{youngtab}
\usepackage{tensor}
\usepackage{slashed}
\usepackage[aligntableaux=center]{ytableau}
\usepackage{url}
\usepackage{bbold}

\usepackage{enumitem} 
\setlist[itemize]{leftmargin=*}
\setlist[enumerate]{leftmargin=*}

\numberwithin{equation}{section}
\allowdisplaybreaks 

\def\<{\langle}
\def\>{\rangle}

\newcommand{\cD}{\ensuremath{\mathcal{D}}}

\newcommand{\cI}{\ensuremath{\mathcal{I}}}

\newcommand{\cO}{\ensuremath{\mathcal{O}}}
\newcommand{\cP}{\ensuremath{\mathcal{P}}}

\newcommand{\ed}{\,.}
\newcommand{\ec}{\,,}
\newcommand{\ecq}{\ec\quad}

\DeclareMathOperator{\trace}{Tr}

\begin{document}


\institution{Weizmann}{Department of Particle Physics and Astrophysics, Weizmann Institute of Science, \cr Rehovot 76100, Israel}

\title{Scalar Correlators in Bosonic Chern-Simons Vector Models}

\authors{Ran Yacoby\worksat{\Weizmann}}

\abstract{We consider the planar limit of Chern-Simons theories coupled to a scalar $\phi$ in the fundamental representation of a $U(N)_k$ gauge group, at both the regular and Wilson-Fisher conformal points. These theories have one single-trace scalar operator $J_0=\bar{\phi}\phi$. We calculate its connected planar $n$-point functions, when all the external momenta are collinear. More specifically, we derive an algebraic recurrence relation that expresses each such $n$-point function in terms of lower-point ones. As an application, we study the four-point function, which was recently shown to be completely fixed up to three truncated solutions to the conformal bootstrap. We show that those truncated solutions do not contribute in our bosonic Chern-Simons theories. The result matches a recent calculation of the four-point function in Chern-Simons theories coupled to a fermion, providing a new test of 3d bosonization duality.
}

\date{}

\maketitle

\tableofcontents
\setlength{\unitlength}{1mm}

\newpage
\section{Introduction}

$U(N)_k$ Chern-Simons (CS) theories coupled to a single scalar or fermion in the fundamental representation have a very simple planar limit ($N,k\to\infty$ with fixed $\lambda=N/k$). In particular, they flow to non-trivial infrared fixed-points, which were shown in \cite{Aharony:2011jz,Giombi:2011kc} to have a higher-spin symmetry at large $N$. This symmetry is generated by an infinite tower of single-trace currents $J_s$ ($s=1,2,\ldots$) of approximately minimal twist $\tau_s=\Delta-s=1+O(1/N)$. Abstract large $N$ conformal field theories (CFTs)  with such a spectrum of single-trace currents were classified in \cite{Maldacena:2012sf}, and are distinguished by the dimension $\Delta_0$ of their single-trace scalar operator $J_0$. They are given by\footnote{There are other related examples of interacting theories with large $N$ higher-spin symmetry. E.g., the theory of $N$ scalars $\vec{\phi}$ with a $(\vec{\phi}\cdot \vec{\phi})^3$ interaction, or generalizations of the above $U(N)_k$ gauge theories to ones with multiple flavors or different gauge groups.}
\begin{itemize}
	\item Regular-Boson/Critical-Fermion: these CFTs have $\Delta_0=1+O(1/N)$ and an interaction $\lambda_6 J_0^3/ N^2$, which is exactly marginal in the planar limit. They can be realized in the $U(N)_k$ scalar theory by tuning its renormalized mass and quartic coupling to zero in the infrared, or by reaching the Gross-Neveu fixed-point in the fermion theory.
	\item Critical-Boson/Regular-Fermion: these CFTs have $\Delta_0=2+O(1/N)$. They can be realized by flowing to a Wilson-Fisher infrared fixed point in the scalar theory, or in the fermionic theory by tuning the renormalized fermion mass to zero in the infrared.
\end{itemize}

The (slightly-broken) higher-spin symmetry is very powerful, leading to constraints that fix the planar three-point functions of all the single-trace operators of the above CFTs in closed form \cite{Maldacena:2012sf}. The analysis of \cite{Maldacena:2012sf} depends only  on the single-trace spectrum, which is identical in the bosonic and femrionic CFTs in each of the cases listed above. It follows that the bosonic and fermoinic CFTs in each of those cases also have identical three-point functions, for some mapping of their parameters ($N$, $\lambda$, and $\lambda_6$ in the regular-boson/critical-fermion case). The  duality map was then found in \cite{Aharony:2012nh,GurAri:2012is} through explicit field theory computations to all orders in $\lambda$. This is the well known example of large $N$ 3d bosonization duality.

The success of \cite{Maldacena:2012sf} suggests that the higher-spin symmetry would also enable fixing all the planar $n\geq 4$-point correlators, in terms of the same parameters used to determine the three-point functions.  Such a result would prove the 3d bosonization duality at large $N$. Nevertheless, achieving this goal using the current methods appears to be difficult. In particular, it is not yet known how to calculate any $n\geq 4$-point function using the approximate higher-spin symmetry constraints, nor is it known whether they are sufficient in order to uniquely fix those correlators.

Recently, \cite{Turiaci:2018nua,Aharony:2018npf} used different principles, such as crossing symmetry and boundedness of correlators in the Regge limit, to constrain the planar four-point function $\langle J_0 J_0 J_0 J_0 \rangle$. It was found to be completely fixed up to three crossing-invariant terms that only involve exchanges of operators with  spins $\leq 2$ in the $J_0\times J_0$ OPE; such truncated solutions to crossing were first constructed in \cite{Heemskerk:2009pn}, and correspond to contact interactions in an $AdS_4$ gravity dual. Furthermore, in the fermionic theories, \cite{Bedhotiya:2015uga,Turiaci:2018nua} calculated the scalar four-point function explicitly, by summing over all planar diagrams in a special kinematic regime in which two of the external momenta are set to zero.\footnote{ The more recent computation of \cite{Turiaci:2018nua} fixes an apparent mistake in the earlier calculation of \cite{Bedhotiya:2015uga}.} The explicit results show that none of the truncated solutions contribute and fixes the four-point function in the fermionic CS theories entirely. In this paper we will preform a similar explicit computation in the bosonic theories, finding an agreement with the results of \cite{Turiaci:2018nua}; this provides a new test of 3d bosonization. In fact, we will obtain a much stronger result for the bosonic theories, and fix all planar $n$-point functions of $J_0(q_i)$ analytically when all the momenta $q_i$ are collinear (but otherwise general).\footnote{This is more general than the kinematic configuration used in \cite{Turiaci:2018nua} for the four-point function in the fermionic theories, but not general enough to deduce the full momenta dependence without more information.}

As in previous work \cite{Giombi:2011kc,Aharony:2012nh,GurAri:2012is}, our strategy is to sum the perturbative series in light-cone gauge. First, in Section \ref{CSVM} we argue that in this gauge, when $\lambda_6=-8\pi^2\lambda^2$ in the regular-boson theory, only rainbow Feynman diagrams contribute to $J_0$ correlators in the collinear momentum frame.\footnote{We are gratefull to G.~Gur-Ari for providing the proof of this fact.} In Section \ref{recurrence}, the sum over those rainbow diagrams is expressed through Dyson-Schwinger equations for certain 1PI vertices, from which the correlators in the regular-boson theory can be extracted. It is then shown that these equations are solved by an algebraic recurrence relation that expresses an $n$-point function in terms of lower-point ones. This recurrence relation is our final result, and it is summarized in Section \ref{examples}. The generalization to arbitrary values of $\lambda_6$ is straightforward, as its contribution to a planar $n$-point function is also determined by products of lower-point functions in momentum space. Finally, the results for the critical-boson theory can be obtained by a Legendre transform, which is, again, a simple algebraic operation in momentum space. As examples, in Section \ref{examples} we extract the $n\leq 4$-point functions using the recurrence relation. The resulting $n\leq 3$-point functions agree with the known expressions derived by \cite{Aharony:2012nh}, and the four-point function matches the one derived in \cite{Turiaci:2018nua} in the fermionic theories, through the 3d bosonization map.

Let us now briefly discuss some possible future applications of our results. The correlators that we computed may shed light on the space of fixed-points of the regular-boson/critical-fermion theories at large $N$. Specifically, at order $1/N$, these theories have a non-trivial beta-function $\beta_{\lambda_6}$ that was shown in \cite{AJM} to lead to either one or three fixed-points. The values of the planar $n\leq 5$-point functions of $J_0$ and its $1/N$ anomalous dimension $\gamma_0$, are required to determine which of those scenarios is realized. Now that we have fixed the relevant correlators, it would be interesting to understand the constraints on (the yet unknown) $\gamma_0$ in each of the scenarios. 

Another possible application is related to Vasiliev higher-spin gravity in $AdS_4$ \cite{Vasiliev:1992av}, which was conjectured in \cite{Giombi:2011kc,Klebanov:2002ja,Sezgin:2003pt,Giombi:2009wh,Chang:2012kt} to be holographically dual to our theories. The formulation of the bulk theory seems to allow for an infinite number of parity violating couplings, all consistent with the higher-spin symmetry. However, only the first of those was related to the `t~Hooft coupling $\lambda$ of the dual CS-matter CFTs \cite{Chang:2012kt}. Of course, these CFTs have no extra parameters, so the possibility of turning on additional independent couplings in Vasiliev theory is puzzling. It is believed that those additional bulk interactions could only contribute to five-point functions and higher, and our results could therefore shed light on their interpretation.\footnote{See also \cite{Vasiliev:2015mka} for a discussion of singularities caused by the extra parity-violating bulk couplings.} To have a full picture, however, would also require calculating the same higher-point functions in Vasiliev theory, which would be quite a formidable task.

Finally, the main future challenge that motivated this work is to construct a complete solution of CS-vector models in the planar limit, at least at the level of correlators. Our results show that it is indeed possible to fix an infinite family of such correlators for collinear external momenta. It would be interesting to generalize our approach to correlators of the currents $J_s$, and also to ones in the fermionic theories. Ultimately, however, obtaining a full solution of large $N$ CS-vector models
would probably require non-Lagrangian CFT methods, such as higher-spin symmetry, crossing, and boundedness in the Regge limit. One insight that can be drawn from our result is that, somewhat surprisingly, the scalar correlators have no logarithmic dependence on the collinear momenta. We expect that this remains true for general momenta. For example, the four-point function is now known exactly and is rational in momentum space, even though it generally contains logarithms in position space in the regular-boson CFT.\footnote{These logarithms correspond to anomalous dimensions of double-trace operators due to the single-trace exchange $J_0\times J_0\sim J_0$, that can appear with an arbitrary coefficient in the regular-boson CFT \cite{Aharony:2018npf}. } This suggests that the higher-spin symmetry constraints might simplify drastically in momentum space (as also emphasized in \cite{Turiaci:2018nua}). Indeed, while in position space these constraints are integro-differential equations in the correlators, they become algebraic in momentum space, which could explain the absence of logarithms. It would be fascinating to study those constraints in momentum space in detail. 

\section{Bosonic Chern-Simons Vector Models}
\label{CSVM}

Let us start with a brief review of bosonic CS vector models; for more details, the reader is referred to \cite{Aharony:2012nh}. The Euclidean action for the $U(N)_k$ regular-boson theory is\footnote{Our conventions are the same as in \cite{Aharony:2012nh}. The covariant derivative is $D_{\mu} = \partial_{\mu}+A_{\mu}$, where the gauge field $A_{\mu}=T^aA^a_{\mu}$ is defined with anti-hermitian $U(N)$ generators $T^a$, normalized such that $\trace T^a T^b = -\frac{1}{2}\delta^{ab}$. }
\begin{align}
S_{\mathrm{RB}}[A_{\mu},\phi] &= -\frac{ik}{4\pi}\int d^3x\varepsilon^{\mu\nu\rho}\trace\left(A_{\mu}\partial_{\nu}A_{\rho} - \frac{2i}{3}A_{\mu}A_{\nu}A_{\rho}\right)\notag\\
&+ D_{\mu}\bar{\phi}D^{\mu}\phi +\frac{\lambda_6}{6N^2}(\bar{\phi}\phi)^3 \ec \label{SRB}
\end{align}
where we implicitly assume a renormalization scheme that fixes the mass and quartic scalar couplings to zero. We will be interested in connected planar correlators of $J_0=\bar{\phi}\phi$, whose generating functional $W_{\mathrm{RB}}[\sigma]$ is defined by
\begin{align}
e^{-W_{\mathrm{RB}}[\sigma]} = \int \cD A \cD\phi e^{-S_{\mathrm{RB}}[A_{\mu},\phi] - \int d^3x \sigma J_0}\biggr|_{\mathrm{planar}} \ed \label{WRB}
\end{align}
As is well known, planar correlators of the critical-boson theory can be obtained from \eqref{WRB} by making $\sigma$ dynamical. In particular, the generating functional $W_{CB}[\zeta]$ of connected planar correlators of $\sigma$ in the critical model is given by a Legendre transform:
\begin{align}
e^{-W_{CB}[\zeta]} = \int D\sigma e^{-W_{RB}[\sigma] + \int d^3x \zeta \sigma} \ed\label{WCB}
\end{align}
In a canonical normalization, both $W_{RB}$ and $W_{CB}$ are proportional to $N$. Therefore, when $N$ is large, $W_{RB}$ ($W_{CB}$) can be interpreted as the generator of 1PI scalar correlators of the critical (regular) model. In the planar limit the two models are therefore simply related, and we will mostly focus on the regular-boson theory until Section \ref{examples}.

As in \cite{Giombi:2011kc,Aharony:2012nh}, we work in light-cone gauge $A_- = 0$ in which the $A^3$ interaction in \eqref{SRB} vanishes.\footnote{The light-cone plane is taken to be the $x^1$--$x^2$ plane. In particular, a momentum vector $\vec{k}=(k^+,k^-,k^3)$ in the light-cone frame is defined as $k^{\pm}=k_{\mp} = \frac{k^1\pm i k^2}{\sqrt{2}}$.} Moreover, in this gauge the propagators of $A_{\mu}$ and $\phi$ are tree-level exact up to $1/N$ corrections, and are given by
\begin{align}
\langle A_{3}^a(-p) A_{+}^b(q)\rangle &= - \langle A_{+}^a(-p) A_{3}^b(q)\rangle =(2\pi)^3\delta^3(q-p) \frac{4\pi i}{k}\frac{1}{p^+} \ec \\
\langle \bar{\phi}_i(-p) \phi^j(q)\rangle &= (2\pi)^3\delta^3(q-p) \frac{\delta_i^j}{p^2} \ec
\end{align}
where $a,b = 1,\ldots, N^2$ are $U(N)$ adjoint indices, while $i,j=1,\ldots,N$  label its fundamental irrep.

\subsection{Reduction to Rainbow Diagrams}

In light-cone gauge, the perturbative expansion of $J_0$ correlators simplifies if one ignores the contributions of the $\lambda_6$ and the seagull  ($\bar{\phi}A^{\mu}A_{\mu}\phi$) interactions. In particular, it becomes a sum over rainbow Feynman diagrams. In this section, we will see that those interactions can be easily accounted for a posteriori, after calculating the correlator by first ignoring their contributions.

\subsubsection{Contributions of $\lambda_6$}

The contribution of the $\lambda_6$ interaction to scalar correlators is very simple in the planar limit. First, recall that $\lambda_6$ is not a parameter of the critical-boson CFT, because the $J_0^3$ interaction is irrelevant in that theory. More technically, for the critical theory,  \eqref{WRB} and \eqref{WCB} imply that
\begin{align}
e^{-W_{CB}[\zeta]} = \int D\sigma e^{-W_{RB}^{\lambda_6=0}[\sigma] + \int d^3x \zeta\sigma - \frac{\lambda_6}{6N^2}\int d^3x \zeta^3}\ec\label{WCB2}
\end{align}
because the path-integral over $\sigma$ imposes $\zeta = J_0$. It follows that $\lambda_6$ may only contributes to contact-terms in correlators of $\sigma$ in the critical theory.  

To understand the contributions in the regular theory, recall that the 1PI action of the regular-boson is
\begin{align}
S_{\mathrm{RB}}^{\mathrm{1PI}}[\zeta] = W_{\mathrm{CB}}[\zeta] \equiv W_{\mathrm{CB}}^{\lambda_6=0}[\zeta] + \frac{\lambda_6}{6N^2}\int d^3x\zeta^3\ed \label{S1PI}
\end{align}
The functional $W_{\mathrm{CB}}^{\lambda_6=0}$, defined in \eqref{S1PI}, generates the 1PI scalar correlators in the regular-boson theory at $\lambda_6=0$, as can be seen by plugging \eqref{S1PI} on the LHS of \eqref{WCB2}. The connected planar correlators of $J_0$ in the regular theory, are given by tree-level diagrams of $\zeta$ that are constructed using the vertices in $S_{\mathrm{RB}}^{\mathrm{1PI}}[\zeta]$. 

The important point is that at tree-level (equivalently, large $N$), each additional insertion of the cubic $\lambda_6$ interaction increases the number of external legs by at least $1$. Therefore, the contribution of $\lambda_6$ to a tree-level  $n$-point  function of $\zeta$ in the theory \eqref{S1PI}, only involves vertices in $S^{\mathrm{1PI}}_{\mathrm{RB}}$ that are at most $O(\zeta^{n-1})$. These contributions are simply products of $m<n$-point functions in momentum space. We conclude that the $\lambda_6$--dependent piece of the $n$-point function is simply constructible from the $m<n$-point functions. 

In Section \ref{1PIvertices}, our strategy will be to find a recurrence relation that fixes each $n$-point function at a particular value of $\lambda_6$ from the full lower-point ones. There is then no real loss of generality in fixing $\lambda_6$, because, as we have just argued, the full dependence on it is anyway accounted for by lower-point correlators. 

\subsubsection{Eliminating the Seagull Interaction}

Above, we argued that the dependence of planar correlators of $J_0$ on $\lambda_6$ is easily accounted for. Now, let us argue that, for special kinematics, the contributions of the seagull vertex are equivalent to turning on a non-zero $\lambda_6$. Specifically, we will prove that
\begin{align}
\langle J_0(-q_1)\cdots J_0(-q_n) \rangle_{\lambda_6=-8\pi^2\lambda^2} = \langle J_0(-q_1)\cdots J_0(-q_n) \rangle_{\lambda_6=0\,, \mathrm{seagull}=0} \ec \label{noseagull}
\end{align}
if all  $q_i^{\pm}=0$. The notation on the RHS of \eqref{noseagull} means that the correlator there is calculated with the action
\begin{align}
S'_{\mathrm{RB}}[A_{\mu},\phi] &=\int d^3 x\left[\frac{k}{4\pi}A_+^a\partial_-A_3^a + \partial_{\mu}\phi^{\dagger}\partial^{\mu}\phi - A_{\mu}^aJ^{a\mu}\right]\ec\label{SRBp}\\
J_{\mu}^a &\equiv \bar{\phi}T^a\partial_{\mu}\phi - \partial_{\mu}\bar{\phi}T^a\phi \ec
\end{align}
with $A_-=0$.

In light-cone gauge, the component $A_+$ of the gauge field appears linearly in the action \eqref{SRB}. It can therefore be integrated out, resulting in some non-local action for the scalars (see \cite{Jain:2012qi}). That action contains a sextic coupling, which can be written in momentum space as
\begin{align}
\delta S_6 = \int \left(\prod_{i=1}^3\frac{d^3 P_i}{(2\pi)^3}\frac{d^3 k_i}{(2\pi)^3}\right)(2\pi)^3\delta^3(P_1+P_2+P_3)C_2(P_1,P_2,k_1,k_2,k_3) \chi(P_1,k_1)\chi(P_2,k_2)\chi(P_3,k_3)\ec \label{S6}
\end{align}
where $\chi(P,k)\equiv \bar{\phi}(\frac{P}{2}-k)\phi(\frac{P}{2}+k)$, and
\begin{align}
C_2(P_1,P_2,k_1,k_2,k_3) \equiv \frac{4\pi^2\lambda^2}{N^2}\frac{(P_1-P_2+2k_1+2k_2)_-(P_1+2P_2+2k_2+2k_3)_-}{(P_1+P_2+2k_1-2k_2)_-(P_1-2k_2+2k_3)_-} + \frac{\lambda_6}{6N^2} \ed \label{C2}
\end{align}
The $\lambda$--dependent term in \eqref{C2} arises from the seagull vertex. 

In planar diagrams of $J_0$, each of the bi-local fields $\chi(P_i,k_i)$ in \eqref{S6} is connected to a distinct scalar loop. Momentum conservation then implies that the $P_i$ is always constrained to be a linear combination of the external momenta (note that $J_0(P) = \int \frac{d^3 k}{(2\pi)^3} \chi(P,k)$). It follows that if all external momenta are collinear, i.e., $q_i^{\pm}=0$, then we can also set $P_i^{\pm}=0$ in the coefficient $C_2(P_1,P_2,k_1,k_2,k_3)$ of the interaction \eqref{S6}:
\begin{align}
\delta S_6 &\to \int \left(\prod_{i=1}^3\frac{d^3 P_i}{(2\pi)^3}\frac{d^3 k_i}{(2\pi)^3}\right)(2\pi)^3\delta^3(P_1+P_2+P_3)\notag\\
&\times\left(\frac{4\pi^2\lambda^2}{N^2}\frac{(k_1+k_2)_-(k_2+k_3)_-}{(k_1-k_2)_-(k_3-k_2)_-} + \frac{\lambda_6}{6N^2}\right) \chi(P_1,k_1)\chi(P_2,k_2)\chi(P_3,k_3)\notag\\
&= \int \left(\prod_{i=1}^3\frac{d^3 P_i}{(2\pi)^3}\frac{d^3 k_i}{(2\pi)^3}\right)(2\pi)^3\delta^3(P_1+P_2+P_3)\left(\frac{8\pi^2\lambda^2+\lambda_6}{6N^2}\right) \chi(P_1,k_1)\chi(P_2,k_2)\chi(P_3,k_3)\ec \label{simplecubic}
\end{align}
where on the last step we symmetrized under permutations of $(P_i,k_i)$. Remarkably, in the planar limit and collinear frame, the non-local sextic interaction \eqref{S6} arising from the seagull term is equivalent to the local $J_0^3(x)$ interaction \eqref{simplecubic}! Moreover, $\delta S_6|_{\lambda_6=-8\pi^2\lambda^2} = 0$, which proves \eqref{noseagull}.\footnote{In the above derivation we assumed there are no insertions of $A_+$ in the path-integral when integrating it out. This is the case for $J_0$ correlators. When single-trace currents $J_{+\mu_2\cdots\mu_s}$ with a component in the $+$ direction are inserted, one has to be more careful.} The fact that the seagull interaction is equivalent to $\lambda_6$ under those circumstances can be seen in the explicit computations of \cite{Aharony:2012nh}.

\section{Recurrence Relation for Scalar Correlators}
\label{recurrence}

In this section we will derive our main result for scalar correlators. Our strategy will be to consider first the non-gauge invariant 1PI correlation functions 
\begin{align}
\langle J_0(-q_1)\cdots J_0(-q_n) \phi_i(-k)\bar{\phi}^j(r)\rangle^{\mathrm{1PI}}_{\lambda_6=-8\pi^2\lambda^2} \equiv \delta_i^j(2\pi)^3\delta^3(\sum_i q_i+k-r)\cdot V_n(\{q_1,\ldots,q_n\};k) \ed \label{V1PIdef}
\end{align}
The $V_n$ are $n$-point 1PI vertices, which can be used as building blocks for the gauge invariant correlators of $J_0$. We work in light-cone gauge and in the frame $q_i^{\pm}=0$, but $k$ in \eqref{V1PIdef} can be general. Moreover, as indicated in \eqref{V1PIdef}, the $V_n$ will be determined for $\lambda_6=-8\pi^2\lambda^2$, which, as argued in the previous section, is equivalent to summing the perturbative expansion with the simplified action \eqref{SRBp}.\footnote{This was proven in Section \ref{CSVM} for gauge-invariant correlators of $J_0$, but it is easy to see that the same argument holds for the 1PI vertices \eqref{V1PIdef}.} 

Under these conditions, $V_1$ was found in \cite{Aharony:2012nh} and is given by\footnote{Throughout the paper, we use $q$ to denote the third component $q^3$ of an external momentum vector $\vec{q}$. If not indicated otherwise, the external momenta always point in the $3$-direction, so there should be no confusion.}
\begin{align}
V_1(\{q\};k)\equiv\frac{2 e^{-2 i \lambda  \arctan\left(\frac{2 k_s}{q}\right)}}{1+ e^{-2 i \lambda  \arctan\left(\frac{2 \Lambda }{q}\right)}} \ec\label{V1}
\end{align}
where $k_s^2\equiv 2k^+k^- = k_1^2+k_2^2$. In \eqref{V1}, $\Lambda$ is a sharp UV cutoff on the momentum in the radial direction of the light-cone plane. In this section, $\Lambda$ will be kept finite for completeness, even though the scalar correlators we are ultimately interested in have no UV divergences in our scheme. Note also that $\langle J_0(-q)\phi_i(-k)\bar{\phi}^j\rangle$ is independent of the sextic coupling, so \eqref{V1} is actually valid for any value of $\lambda_6$. 

In what follows we will derive a recurrence relation for the $V_n$ defined in \eqref{V1PIdef}, which fixes them in terms of the known $V_1$ given in \eqref{V1}. We will then show that the gauge-invariant correlators of $J_0$ can be very simply inferred from the 1PI vertices \eqref{V1PIdef}.

\subsection{1PI-Vertices}
\label{1PIvertices}

The perturbative expansion of $V_n$ in \eqref{V1PIdef} can be organized efficiently with the aid of Dyson-Schwinger (DS) equations. To construct those equations, we need to set up some notation. Let $\cP(n)\equiv \cP\left(\left\{q_1,\ldots,q_n\right\}\right)$ be the set of all ordered partitions of $\{q_1,\ldots,q_n\}$.\footnote{By ordered partitions we mean all the ways to place brackets on the ordered sequence $\{q_1,\ldots,q_n\}$ such that no bracket is contained in another; i.e., the brackets have the form $(X)(Y)\cdots$, but not $(X(Y)Z)\cdots$. This is a subset of what is usually referred to as non-crossing partitions.} For example,
\begin{align*}
\cP(3) = \{ \{ \{q_1\}\ec\{q_2\}\ec\{q_3\} \} \ec \{ \{q_1\ec q_2\}\ec\{q_3\}\} \ec \{ \{q_1\} \ec \{q_2\ec q_3\} \} \ec \{ \{q_1\ec q_2\ec q_3\}\}\} \ed
\end{align*}
It is convenient to treat such partitions as arrays and use index notation. For example, $p_i$ denotes the $i$-th set within $p$, and $p_{ij}$ the $j$-th element of the $i$-th set. For $p\in \cP(n)$ we define $\bar{p}_i\equiv \sum_{j=1}^{|p_i|} p_{ij}$, and $\hat{p}_i\equiv \bar{p}_1+\cdots+\bar{p}_i$, where $|X|$ is the number of elements of the set $X$. For example, if $p = \{ \{q_1\ec q_2\}\ec\{q_3\}\} \in \cP(3)$, then
\begin{gather*}
\bar{p} = \{ q_1+q_2 \ec q_3 \} \ecq \hat{p} = \{ q_1+q_2 \ec q_1+q_2+q_3 \} \ec \\
|p| = |\bar{p}|=|\hat{p}|=|p_1|=2\ecq |p_2| = 1\ed
\end{gather*}
In what follows these definitions will be used extensively, and the letter $p$ will always be reserved to denote such partitions, which depend on the external momenta $q_i$. 

\begin{figure}[t!]
	\centering
	\includegraphics[width=0.8\textwidth]{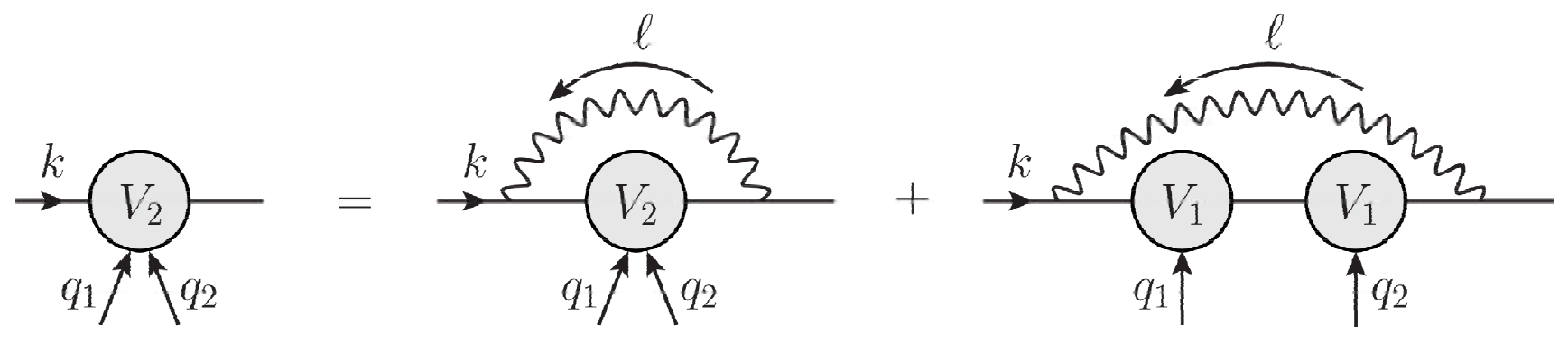}
	\caption{The DS equations of the planar 1PI vertex $V_2$. The solid and wiggly lines are scalar and gluong propagators, respectively.}
	\label{fig:DSV2}
\end{figure}

Let us now construct the DS equations. We first implicitly only consider contributions to $V_n$ from planar diagrams in which the $J_0(-q_i)$ insertions in \eqref{V1PIdef} are ordered as $(q_1,\ldots,q_n)$ along the (single) scalar line. At the end one has to sum over permutations of $q_i$ to obtain the full answer. Up to those permutations, the DS equations for the vertices \eqref{V1PIdef} can be written as
\begin{align}
V_n(\{q_1,\ldots,q_n\};k) &= \delta_{n,1} + 4\pi i \lambda (q_1+\cdots+q_n) \!\! \sum_{p\in \cP(n)} \int \!\frac{d^3\ell}{(2\pi)^3} \frac{(\ell+k)_+}{(\ell-k)_+} \frac{1}{\ell^2} \prod_{i=1}^{|p|} \frac{V_{|p_i|}(p_i;\ell)}{(\ell+\hat{p}_i)^2}\ed \label{DS}
\end{align}
Each of the terms in the sum in \eqref{DS} represents one diagram in the DS equation. The $\frac{(\ell+k)_+}{(\ell-k)_+}$ factor comes from the gauge field line that encloses a product of lower vertices. See Figure \ref{fig:DSV2} for an example. 

The first important observation is that the RHS of \eqref{DS} is independent of $k^3$, and therefore, so is $V_n(\{q_1,\ldots,q_n\};k)$ on the LHS. Furthermore, because our gauge preserves rotational invariance in the light-cone plane, $V_n$ can only depend on $k_s^2 = 2k^+k^- = k_1^2+k_2^2$; i.e., $V_n(\{q_1,\ldots,q_n\};k)=V_n(\{q_1,\ldots,q_n\};k_s)$. It follows that when all the $q_i$ (and therefore $\hat{p}_i$) point in the $3$-direction,  the only dependence on the polar angle in the light-cone $\ell$-plane is through the $\frac{(\ell+k)_+}{(\ell-k)_+}$ factor in the integrand on the RHS of \eqref{DS}. Integrating over this angular variable results in\footnote{We used $\int_0^{2\pi} d\theta_{\ell}\frac{(\ell+k)_+}{(\ell-k)_+} = 2\pi(2\Theta(\ell_s-k_s)-1)$, and $d^3\ell = \ell_s d\ell_s d\theta_{\ell}d\ell_3$, where $\ell^{\pm} = \frac{\ell_s}{\sqrt{2}} e^{\pm\theta_{\ell}}$.}
\begin{align}
V_n(\{q_1,\ldots,q_n\};k_s) &= \delta_{n,1} + 2 i \lambda \left(\sum_{i=1}^n q_i\right) \!\!\! \sum_{p\in \cP(n)}\! \int_0^{\Lambda} \!\!\!d\ell_s \left( 2\Theta\left(\ell_s-k_s\right) - 1 \right) \!\int_{-\infty}^{\infty}\! \frac{d\ell_3}{2\pi}\frac{\ell_s}{\ell_3^2+\ell_s^2} \prod_{i=1}^{|p|} \frac{V_{|p_i|}(p_i;\ell_s)}{(\ell_3+\hat{p}_i)^2+\ell_s^2}\ec \label{intDS}
\end{align}
where $\Theta$ is the Heaviside step function.

By taking a derivative with respect to $k_s$ in \eqref{intDS}, one obtains the differential DS equation
\begin{align}
\frac{dV_n(\{q_1,\ldots,q_n\};k_s)}{dk_s} = -4 i \lambda \left(\sum_{i=1}^nq_i\right) \!\! \sum_{p\in \cP(n)} \int_{-\infty}^{\infty}\! \frac{d\ell_3}{2\pi}\frac{k_s}{\ell_3^2+k_s^2} \prod_{i=1}^{|p|} \frac{V_{|p_i|}(p_i;k_s)}{(\ell_3+\hat{p}_i)^2+k_s^2}\ed \label{diffDS}
\end{align}
The RHS of this expression contains terms that are of the same form as ones which already appear in DS equations of $V_{m<n}$. For instance, isolating the contribution of the single partition of length $1$ gives
\begin{align}
\frac{dV_n(\{q_1,\ldots,q_n\};k_s)}{dk_s}\subset -4 i \lambda \left(\sum_{i=1}^nq_i\right)\int_{-\infty}^{\infty} \frac{d\ell_3}{2\pi} \frac{k_s V_n(\{q_1,\ldots,q_n\};k_s)}{(\ell_3^2+k_s^2)((\ell_3+q_1+\ldots +q_n)^2+k_s^2)} \ed \label{onepartition}
\end{align}
The RHS of \eqref{onepartition} is of the same form as the one in the DS equation of $V_1$, if we identify $V_n(\{q_1,\ldots,q_n\};k_s)\to b_n V_1(\{q_1+\cdots +q_n\};k_s)$. We therefore expect that $V_n = b_n V_1 + \cdots$. More generally, the structure of \eqref{diffDS} suggests that $V_n$ could be expressed in terms of lower-point vertices. After some experimentation with low $n$'s one is led to the following  ansatz,
\begin{align}
V_n(\{q_1,\ldots,q_n\};k_s) &= a_n(\{q_1,\ldots,q_n\}) V_1(\{q_1\};k_s)\cdots V_1(\{q_n\};k_s) + \notag\\
&~ \sum_{p\in \cP'(n)} \left[(-1)^{|p|+1}\left( \prod_{i=1}^{|p|}b_{|p_i|}(p_i)\right)V_{|p|}(\bar{p};k_s)\right] \ec \label{ansatz}
\end{align}
where $\cP'(n) = \cP(n) - \{\{q_1\},\ldots,\{q_n\}\}$, is the set of ordered partitions of length $<n$, and the coefficients $a_n$ and $b_n$ are some functions of the $q_i$ (and independent of $k_s$).

Let us now discuss how to determine the coefficients in \eqref{ansatz}. The $a_i$ are determined by the differential equation recursively. To see this we plug the ansatz \eqref{ansatz} into \eqref{diffDS} and equate the terms proportional to $V_1(\{q_1\};k_s)\cdots V_1(\{q_n\};k_s)$:
\begin{gather}
a_n(\{q_1,\ldots,q_n\}) \frac{d}{dk_s}\left(V_1(\{q_1\};k_s)\cdots V_1(\{q_n\};k_s)\right)  \notag\\
=-4 i \lambda (q_1+\cdots+q_n) V_1(\{q_1\};k_s)\cdots V_1(\{q_n\};k_s) \!\! \sum_{p\in \cP(n)} \int_{-\infty}^{\infty}\! \frac{d\ell_3}{2\pi}\frac{k_s}{\ell_3^2+k_s^2} \prod_{i=1}^{|p|} \frac{a_{|p_i|}(p_i)}{(l_3+\hat{p}_i)^2+k_s^2} \ed
\end{gather}
Plugging in $V_1$, which was defined in \eqref{V1}, leads to
\begin{gather}
a_n(\{q_1,\ldots,q_n\})\left[\frac{q_1}{q_1^2+4k_s^2}+\cdots+\frac{q_n}{q_n^2+4k_s^2}\right] \notag \\
=(q_1+\cdots+q_n) \!\! \sum_{p\in \cP(n)} \int_{-\infty}^{\infty}\! \frac{d\ell_3}{2\pi}\frac{k_s}{\ell_3^2+k_s^2} \prod_{i=1}^{|p|} \frac{a_{|p_i|}(p_i)}{(\ell_3+\hat{p}_i)^2+k_s^2}\label{aEq}
\end{gather}
The integrals appearing in \eqref{aEq} can be solved using contour integration:
\begin{align}
\cI(p;k_s) \equiv \int_{-\infty}^{\infty} \frac{k_s\, d\ell_3}{2\pi} \prod_{i=0}^{|p|}\frac{1}{(\ell_3+\hat{p}_i)^2+k_s^2} = \sum_{\substack{i,j=0\\i<j}}^{|p|}\left[\left(\prod_{\substack{m=0\\m\neq i,j}}^{|p|} \frac{1}{\hat{p}_{mi}\hat{p}_{mj}}\right)\frac{1}{\hat{p}_{ij}^2+4k_s^2}\right]\ec \label{int}
\end{align}
where $\hat{p}_{ij}\equiv \hat{p}_i-\hat{p}_j$, and $\hat{p}_0\equiv 0$.  Plugging \eqref{int} into \eqref{aEq}, one can determine $a_n$ recursively, resulting in
\begin{align}
a_n(\{q_1,\ldots,q_n\}) = \frac{q_1+\cdots+q_n}{q_1q_2\cdots q_n (q_1+q_2)(q_2+q_3)\cdots(q_{n-1}+q_n)} \ed \label{aSol}
\end{align}
Note that $a_1=1$.

The coefficients $b_n$ in \eqref{ansatz} are determined from the integral equation \eqref{intDS} in the following way. Plugging the differential DS equation \eqref{diffDS} into the integral one \eqref{intDS} we obtain the relation
\begin{gather}
V_n(\{q_1,\ldots,q_n\};k_s) = \delta_{n,1} -\frac{1}{2}\int_0^{\Lambda}\left(2\Theta(\ell_s-k_s)-1\right) \frac{dV_n(\{q_1,\ldots,q_n\};\ell_s)}{d\ell_s}\notag\\
\Downarrow\notag\\
V_n(\{q_1,\ldots,q_n\};0) + V_n(\{q_1,\ldots,q_n\};\Lambda)= 2\delta_{n,1} \ed \label{intCond}
\end{gather}
Assuming by induction that \eqref{intCond} is satisfied by $V_{i<n}$, and plugging this condition in the recurrence relation \eqref{ansatz}, leads to
\begin{align}
b_n &= -\frac{a_n}{2} \left(V_1(\{q_1\};\Lambda)\cdots V_1(\{q_n\};\Lambda)+V_1(\{q_1\};0)\cdots V_1(\{q_n\};0)\right) \notag\\
&= -\,2^{n-1}\,a_n\left(\prod_{i=1}^{n}\frac{1}{1+e^{-2i\lambda\arctan\left(2\frac{\Lambda}{q_i}\right)}}\right) \left(1 + e^{-2i\lambda\sum_{i=1}^{n}\arctan\left(2\frac{\Lambda}{q_i}\right)}\right) \ed \label{bSol}
\end{align}
Note that $b_1=-a_1=-1$. In Appendix \ref{proof} we will prove that \eqref{ansatz} indeed solves the DS equation \eqref{DS}, with $a_n$ and $b_n$ given in \eqref{aSol} and \eqref{bSol}, respectively. 

\subsection{Gauge-Invariant Correlators}

\begin{figure}[t!]
	\centering
	\includegraphics[width=0.6\textwidth]{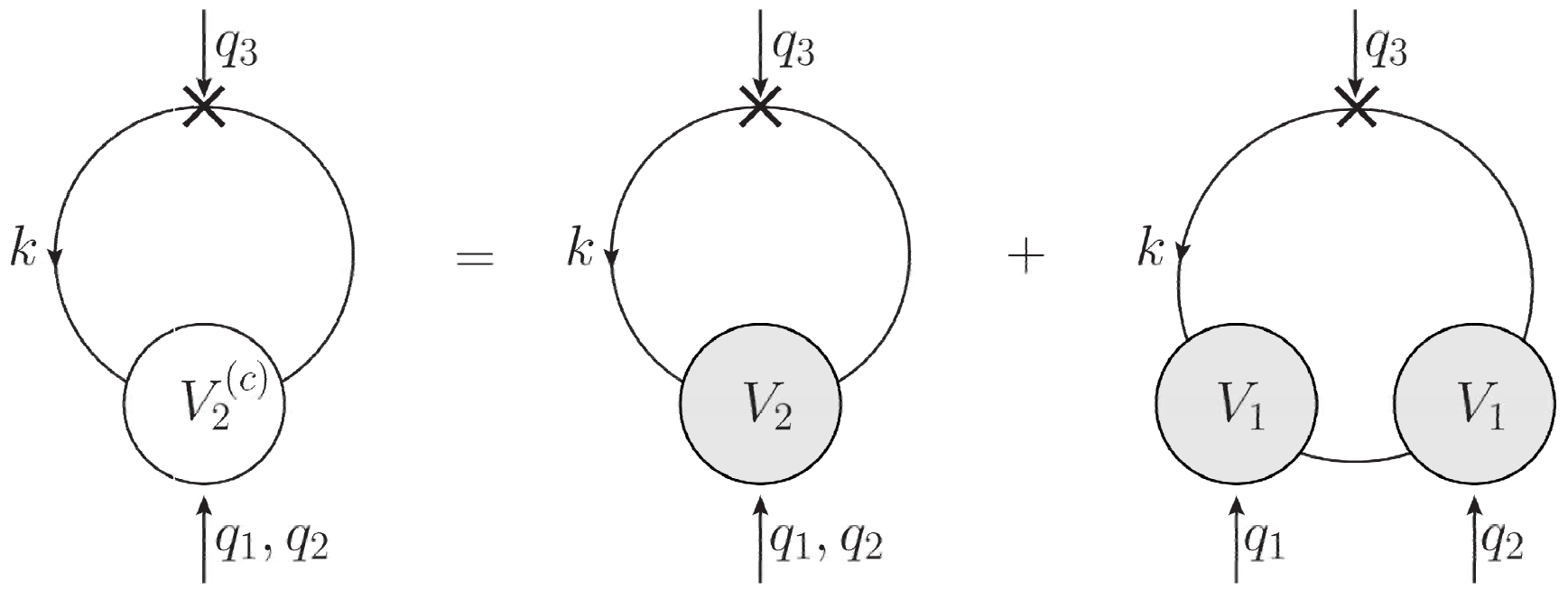}
	\caption{The connected scalar three-point function in terms of the connected $V^{(c)}_2$ vertex, and the 1PI vertices $V_{1,2}$, up to permutations}
	\label{fig:G3ord}
\end{figure}

Let us now discuss how the connected planar $n$-point function
\begin{align}
\langle J_0(-q_1)\cdots J_0(-q_n)\rangle_{\lambda_6=-8\pi^2\lambda^2}^{\mathrm{conn.}} \equiv (2\pi)^3\delta^3(\sum_iq_i) \cdot G^{(n)}_{\lambda_6=-8\pi^2\lambda^2}(q_1,\ldots,q_{n-1})\ec \label{npnt}
\end{align}
can be extracted from the 1PI vertices \eqref{V1PIdef}. Consider first the contributions $G^{(n)}_{\mathrm{ord.}}$ to \eqref{npnt}, that come from diagrams in which the insertions are ordered along the scalar loop. These contributions can be constructed from the connected and amputated vertex function $V_n^{(c)}$ with ordered insertions:
\begin{align}
\langle J_0(-q_1)\cdots J_0(-q_n) \phi_i(-k)\bar{\phi}^j(r)\rangle^{\mathrm{conn.-amp.- ord.}}_{\lambda_6=-8\pi^2\lambda^2} \equiv \delta_i^j(2\pi)^3\delta^3(\sum_i q_i+k-r)\cdot V_n^{(c)}(\{q_1,\ldots,q_n\};k) \ed\label{Vnc}
\end{align}
The $V_n^{(c)}$ can be written in terms of the 1PI vertices as
\begin{align}
V_n^{(c)}(\{q_1,\ldots,q_n\};k) = \sum_{p\in \cP(n)} V_{|p_{|p|}|}(p_{|p|};k_s) \prod_{i=1}^{|p|-1}\frac{V_{|p_i|}(p_i;k_s)}{(k+\hat{p}_i)^2} \ed
\end{align}

The correlator $G^{(n)}_{\mathrm{ord.}}$ is obtained from $V_{n-1}^{(c)}$ by closing the scalar line on a new $J_0$ insertion (see Figure \ref{fig:G3ord} for an example). This gives
\begin{align}
G^{(n)}_{\mathrm{ord.}}(q_1,\ldots,q_{n-1}) = N\int \frac{d^3k}{(2\pi)^3} \frac{V_{n-1}^{(c)}(\{q_1,\ldots,q_{n-1}\};k)}{k^2(k+q_1+\cdots+q_{n-1})^2}= N\sum_{p\in \cP(n-1)} \int \frac{k_s dk_s dk^3}{(2\pi)^2} \frac{1}{k^2}\prod_{i=1}^{|p|} \frac{V_{|p_i|}(p_i;k_s)}{(k+\hat{p}_i)^2}\ec \label{Gnint}
\end{align}
where on the second equality we preformed the trivial integral over the angular direction in the light-cone $k$-plane. Comparing \eqref{Gnint} to \eqref{diffDS} we find
\begin{align}
G^{(n)}_{\mathrm{ord.}}(q_1,\ldots,q_{n-1}) &= \frac{N i}{8\pi\lambda}\frac{1}{q_1+\cdots+q_{n-1}} \int_0^{\Lambda} d_{k_s}V_{n-1}(\{q_1,\ldots,q_{n-1}\};k_s)\notag\\ 
&= \frac{N i}{8\pi\lambda}\frac{1}{q_1+\cdots+q_{n-1}}\left(V_{n-1}(\{q_1,\ldots,q_{n-1}\};\Lambda)-V_{n-1}(\{q_1,\ldots,q_{n-1}\};0)\right) \ed
\end{align}
Finally, using \eqref{intCond} we arrive at the following simple solution, for the correlators of $J_0$ in terms of the 1PI vertices:
\begin{align}
G^{(n)}_{\mathrm{ord.}}(q_1,\ldots,q_{n-1}) = -\frac{N i}{4\pi\lambda}\frac{1}{q_1+\cdots+q_{n-1}} (V_{n-1}(\{q_1,\ldots,q_{n-1}\}; k_s=0)-\delta_{n,2}) \ed
\end{align}
The full correlator \eqref{npnt} is obtained by summing over permutations:
\begin{align}
G^{(n)}_{\lambda_6=-8\pi^2\lambda^2}(q_1,\ldots,q_{n-1}) \equiv \sum_{\sigma\in S_{n-1}} G^{(n)}_{\mathrm{ord.}}(q_{\sigma(1)},\ldots,q_{\sigma(n-1)})\ed
\end{align}

Until now we have fixed the sextic scalar coupling to $\lambda_6=-8\pi^2\lambda^2$. To get the final answer we have to add the contributions from more general values of $\lambda_6$. As discussed in Section \ref{CSVM}, this can also be done recursively, as will be seen in examples bellow.

\section{Examples}
\label{examples}

In this section we will use our recurrence relation \eqref{ansatz} to derive the planar $n\leq 4$-point functions of $J_0$ in the regular and critical bosonic CS theories. Let us first summarize our prescription. Up to permutations of the external momenta, the 1PI vertices are given by the recurrence relation
\begin{align}
V_n(\{q_1,\ldots,q_n\};k_s) &= a_n(\{q_1,\ldots,q_n\}) V_1(\{q_1\};k_s)\cdots V_1(\{q_n\};k_s) + \notag\\
&~ \sum_{p\in P'(n)} \left[(-1)^{|p|+1}\left( \prod_{i=1}^{|p|}b_{|p_i|}(p_i)\right)V_{|p|}(\bar{p};k_s)\right] \ec \label{V1PI}
\end{align}
where
\begin{align}
a_n(\{q_1,\ldots,q_n\}) &= \frac{q_1+\cdots+q_n}{q_1q_2\cdots q_n (q_1+q_2)(q_2+q_3)\cdots(q_{n-1}+q_n)} \ec \label{asol}\\
b_n(\{q_1,\ldots,q_n\}) &= -\,2^{n-1}\,a_n(\{q_1,\ldots,q_n\})\left(\prod_{i=1}^{n}\frac{1}{1+e^{-\pi i \lambda\sgn(q_i)}}\right) \left(1 + e^{-\pi i\lambda\sum_{i=1}^{n}\sgn(q_i)}\right)\ed \label{bsol}
\end{align}
The initial condition for the recurrence relation \eqref{V1PI} is provided by the 1-point vertex \eqref{V1},
\begin{align}
V_1(\{q\};k_s) = \frac{2e^{-2 i \lambda \arctan\left(2\frac{k_s}{q}\right)}}{1+e^{-\pi i \lambda\sgn(q)}} \ed \label{V1nocut}
\end{align}
Equations \eqref{bsol} and \eqref{V1nocut} were obtained by taking $\Lambda\to\infty$ in \eqref{bSol} and \eqref{V1}. There is no obstruction to remove the cutoff in our calculations. 

With the above definitions, the connected $n$-point functions of $J_0$ at $\lambda_6=-8\pi^2\lambda^2$ are given by
\begin{align}
G^{(n)}_{\lambda_6=-8\pi^2\lambda^2}(q_1,\ldots,q_{n-1}) = \sum_{\sigma\in S_{n-1}} G^{(n)}_{\mathrm{ord.}}(q_{\sigma(1)},\ldots,q_{\sigma(n-1)})\ec\label{Gfull}
\end{align}
where
\begin{align}
G^{(n)}_{\mathrm{ord.}}(q_1,\ldots,q_{n-1}) = -N\frac{i}{4\pi\lambda}\frac{1}{q_1+\cdots+q_{n-1}} (V_{n-1}(\{q_1,\ldots,q_{n-1}\}; k_s=0)-\delta_{n,2}) \ed\label{Gord}
\end{align}
To get the full answer for any $\lambda_6$, let $G^{(n)}_6$ denote the full $\lambda_6$--dependent contributions to the $n$-point function. Then
\begin{align}
G^{(n)}(q_1,\ldots,q_{n-1}) = G^{(n)}_{\lambda_6=-8\pi^2\lambda^2}(q_1,\ldots,q_{n-1}) - G^{(n)}_6(q_1,\ldots,q_{n-1})\bigr|_{\lambda_6=-8\pi^2\lambda^2} + G^{(n)}_6(q_1,\ldots,q_{n-1}) \ed \label{Ggeneral}
\end{align}
As discussed above, $G^{(n)}_6$ is determined from the $G^{(i)}$ with $i<n$. Let us now proceed with a few examples.

\subsection{Two-Point}

In the two-point function, there are no contributions from $\lambda_6$. We find
\begin{align}
G^{(2)}(q) = -\frac{Ni}{4\pi\lambda}\frac{1}{q}(V_1(\{q\};0)-1) = N\frac{\tan\left(\frac{\pi\lambda}{2}\right)}{4\pi\lambda}\frac{1}{|q|}\ec \label{2pnt}
\end{align}
which is indeed the answer derived in \cite{Aharony:2012nh}. The result in the critical theory, obtained by a Legendre transform of \eqref{2pnt}, is simply minus the inverse of \eqref{2pnt}.

\subsection{Three-Point}

\begin{figure}[t!]
	\centering
	\includegraphics[width=0.2\textwidth]{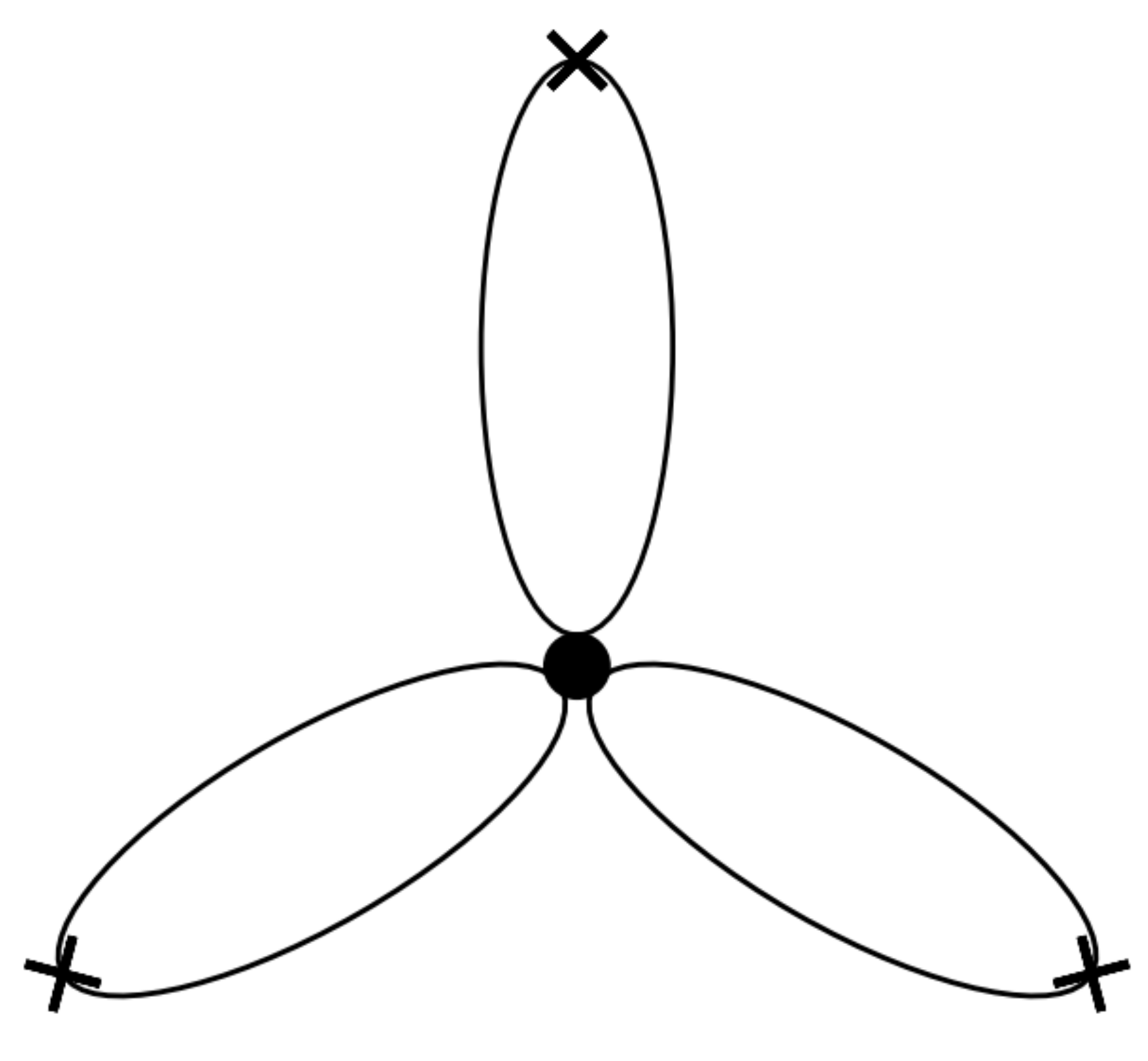}
	\caption{Contribution of the $(\bar{\phi}\phi)^3$ interaction to the planar three-point function of $J_0=\bar{\phi}\phi$. The sum over planar gauge interactions within each of the scalar loops is suppressed.}
	\label{fig:threepoint6}
\end{figure}

For the three-point function there is an $O(\lambda_6)$ contribution (see Figure \ref{fig:threepoint6}). It can be expressed as a product of the two-point functions, which were already determined in \eqref{2pnt}:
\begin{align}
G^{(3)}_{6}(q_1,q_2) &= -\frac{\lambda_6}{N^2}G^{(2)}(q_1)G^{(2)}(q_2)G^{(2)}(-q_1-q_2)\notag\\
&=- N \lambda_6 \left(\frac{\tan\left(\frac{\pi\lambda}{2}\right)}{4\pi\lambda}\right)^3\frac{1}{|q_1||q_2||q_1+q_2|} \ed\label{3pntLam6}
\end{align}
Moreover, using our recurrence relation we find
\begin{align}
G^{(n)}_{\lambda_6=-8\pi^2\lambda^2}(q_1,q_2) &= -N\frac{i}{4\pi\lambda} \frac{1}{q_1+q_2}\left(V_2(\{q_1,q_2\};0) + V_2(\{q_2,q_1\};0)\right) \notag\\
&= \frac{N}{2\pi\lambda}\frac{\tan\left(\frac{\pi\lambda}{2}\right)}{\cos^2\left(\frac{\pi\lambda}{2}\right)} \frac{1}{|q_1||q_2||q_1+q_2|} \ed\label{G3specific}
\end{align}
The correlator for general values of $\lambda_6$ can be obtained by plugging \eqref{3pntLam6} and \eqref{G3specific} into \eqref{Ggeneral}, resulting in
\begin{align}
G^{(3)}(q_1,q_2) = \frac{N}{2\pi\lambda}\left[\frac{\tan\left(\frac{\pi\lambda}{2}\right)}{\cos^2\left(\frac{\pi\lambda}{2}\right)} - \frac{1}{4} \tan^3\left(\frac{\pi\lambda}{2}\right)\left(1+\frac{\lambda_6}{8\pi^2\lambda^2}\right)\right]\frac{1}{|q_1||q_2||q_1+q_2|} \ed \label{3pnt}
\end{align}
The result \eqref{3pnt} is precisely the three-point function found in \cite{Aharony:2012nh}. The Legendre transform of \eqref{3pnt} is a pure contact-term, so the three-point function in the critical-boson theory vanishes at separated points.

\subsection{Four-Point}

\begin{figure}[t!]
	\centering
	\includegraphics[width=0.4\textwidth]{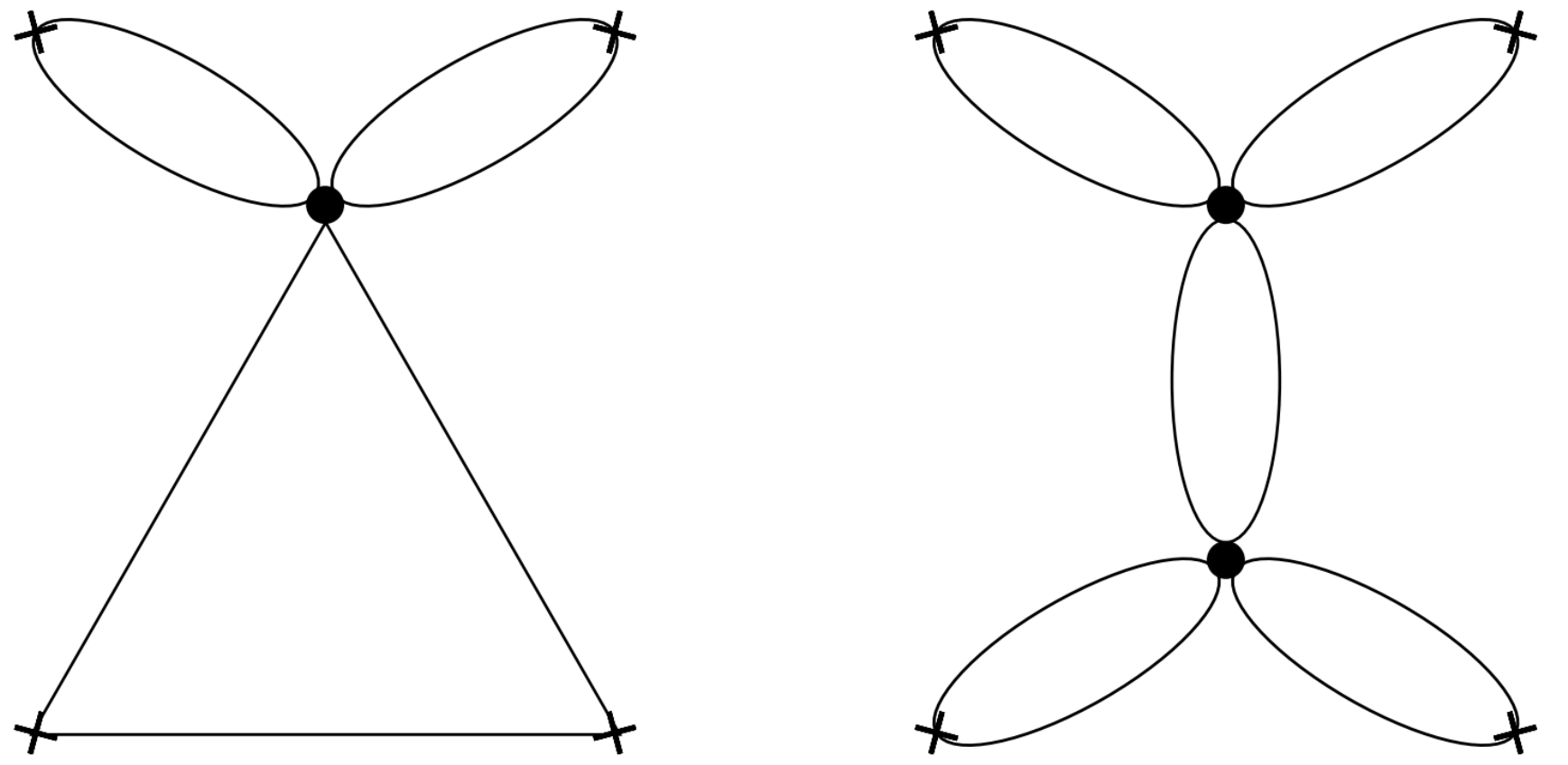}
	\caption{Contribution of the $(\bar{\phi}\phi)^3$ interaction to the planar four-point function of $J_0=\bar{\phi}\phi$ (up to permutations). The sum over planar gauge interactions within each of the scalar loops is suppressed.}
	\label{fig:fourpoint6}
\end{figure}

The four-point function can be simply expressed using the one of the free scalar theory, which, in turn, can be written in terms of the massless box integral\footnote{See \cite{Anninos:2017eib} for a recent calculation of the integral \eqref{box}.}
\begin{gather}
I_4(q_1,q_2,q_3) = \int\frac{d^3k}{(2\pi)^3}\frac{1}{k^2(k+q_1)^2(k+q_1+q_2)^2(k+q_1+q_2+q_3)^2} \notag\\
=\frac{|q_3|\left(|q_2| | q_2+q_3| + | q_1+q_2| | q_1+q_2+q_3| \right)+ | q_1|  \left(| q_2| | q_1+q_2| + | q_2+q_3|  | q_1+q_2+q_3| \right)}{8 | q_1|  | q_2|  | q_1+q_2|  | q_3|  | q_2+q_3|  | q_1+q_2+q_3|  \left(| q_1|  | q_3| +| q_1+q_2|  | q_2+q_3| +| q_2|  | q_1+q_2+q_3| \right)} \ed \label{box}
\end{gather}
Note that \eqref{box} is true for general momenta $q_i$, and not just in the collinear frame. For a single free complex scalar we have
\begin{align}
G^{(4)}_{\text{complex-bos.}}(q_1,q_2,q_3) = 2(I_4(q_1,q_2,q_3) + (q_1\leftrightarrow q_3) + (q_2\leftrightarrow q_3)) \ed \label{G4free}
\end{align}

In the regular-boson CS theory, the four-point function has $O(\lambda_6)$ and $O(\lambda_6^2)$ contributions (see Figure \ref{fig:fourpoint6}). These contributions can be written as products of the two-point and three-point functions that we have already determined in \eqref{2pnt} and \eqref{3pnt}, and are given by
\begin{align}
G^{(4)}_6(q_1,q_2,q_3) &=-N\frac{\lambda_6}{\pi\lambda}\left(\frac{\tan\left(\frac{\pi\lambda}{2}\right)}{4\pi\lambda}\right)^2\left(\frac{\tan\left(\frac{\pi\lambda}{2}\right)}{\cos^2\left(\frac{\pi\lambda}{2}\right)}-\frac{1}{4}\tan^3\left(\frac{\pi\lambda}{2}\right)\left(1+\frac{\lambda_6}{16\pi^2\lambda^2}\right) \right)\notag\\
&\times \left(\frac{1}{|q_1||q_2||q_3||q_1+q_2||q_1+q_2+q_3|} + (q_1\leftrightarrow q_3) + (q_2\leftrightarrow q_3)\right)\ed \label{4pntLam6}
\end{align}
The result of the recurrence relation can be written in terms of the free scalar four-point function \eqref{G4free} as
\begin{gather}
G^{(4)}_{\lambda_6=-8\pi^2\lambda^2}(q_1,q_2,q_3) = -N \frac{i}{4\pi\lambda}\frac{1}{q_1+q_2+q_3}\sum_{\sigma\in S_3}V_3(\{q_{\sigma(1)},q_{\sigma(2)},q_{\sigma(3)}\};0) \notag\\
= N\frac{2}{\pi\lambda}\frac{\tan\frac{\pi\lambda}{2}}{\cos^2\frac{\pi\lambda}{2}}G^{(4)}_{\text{complex-bos.}} + N\frac{1}{\pi\lambda}\frac{\tan^3\frac{\pi\lambda}{2}}{\cos^2\frac{\pi\lambda}{2}}\left(\frac{1}{|q_1||q_2||q_3||q_1+q_2||q_1+q_2+q_3|}+(q_1\leftrightarrow q_2)+ (q_2\leftrightarrow q_3)\right) \ed \label{4pntrec}
\end{gather}
Plugging \eqref{4pntLam6} and \eqref{4pntrec} into \eqref{Ggeneral}, the full four-point function in the regular theory is found to be
\begin{align}
G^{(4)}(q_1,q_2,q_3) &= N\frac{2}{\pi\lambda}\frac{\tan\left(\frac{\pi\lambda}{2}\right)}{\cos^2\left(\frac{\pi\lambda}{2}\right)} G^{(4)}_{\text{complex-bos.}}(q_1,q_2,q_3) +G^{(4)}_{\lambda_6}(q_1,q_2,q_3) \notag\\
&+ \frac{N(17-\cos\pi\lambda)\tan^3\frac{\pi\lambda}{2}}{32\pi\lambda\cos^2\frac{\pi\lambda}{2}}\left(\frac{1}{|q_1||q_2||q_3||q_1+q_2||q_1+q_2+q_3|}+(q_1\leftrightarrow q_3)+(q_2\leftrightarrow q_3)\right)\ed\label{4pnt}
\end{align}

Let us now analyze \eqref{4pnt}. One consistency check is that in the $\lambda\to 0$ limit, \eqref{4pnt} correctly reproduces the answer for $N$  complex scalars with a $(\bar{\phi}\phi)^3$ interaction. Moreover, normalizing $J_0$ such that its two-point function matches the free scalar one, we find that the $\lambda$-dependent pre-factor of the first term on the RHS of \eqref{4pnt} becomes
\begin{align}
N\frac{2}{\pi\lambda}\frac{\tan\left(\frac{\pi\lambda}{2}\right)}{\cos^2\left(\frac{\pi\lambda}{2}\right)}\left(2N\frac{\tan\left(\frac{\pi\lambda}{2}\right)}{\pi\lambda}\right)^{-2} = \frac{\pi\lambda}{N\sin\pi\lambda} = \frac{2}{c_T(\lambda)} \ed
\end{align}
Above, $c_T$ is the coefficient of the two-point function of the stress tensor, which was determined in \cite{Aharony:2012nh}. As argued in \cite{Turiaci:2018nua,Aharony:2018npf}, this is the expected $\lambda$--dependence of the free field theory term in \eqref{4pnt}.

The last two terms in \eqref{4pnt} correspond to tree-level four-point Witten diagrams in $AdS_4$ constructed with a $\Phi^3$ interaction, where $\Phi$ is the bulk field dual to $J_0$.\footnote{In position space $G^{(4)}_{\lambda_6}$ can be written in terms of $\bar{D}$-functions.} These two terms cancel upon setting
\begin{align}
\lambda_6 = -\frac{4 \pi ^2 \lambda ^2}{\cos^2\left(\frac{\pi  \lambda }{4}\right)}  \left(\cos \left(\frac{\pi  \lambda }{2}\right)-3\right) \ed \label{lam6free}
\end{align}
This is as expected. Indeed, from the analysis \cite{Maldacena:2012sf} of Maldacena and Zhiboedov we have (below, $s\neq 0$ and $\langle\cdot\rangle_{\mathrm{bos.}}$ is the correlator of a free real scalar):
\begin{gather}
\langle J_s J_s\rangle = \tilde{N}\langle\cdot\rangle_{\mathrm{bos.}} \ecq \langle J_0 J_0 \rangle = \frac{\tilde{N}}{1+\tilde{\lambda}^2}\langle\cdot\rangle_{\mathrm{bos.}} \ec\\
\langle J_0 J_0 J_s \rangle = \tilde{N}\frac{1}{1+\tilde{\lambda}^2}\langle\cdot \rangle_{\mathrm{bos.}} \ecq \langle J_0 J_0 J_0 \rangle = \tilde{N}\frac{1}{(1+\tilde{\lambda}^2)^2}\left(1 + \frac{\tilde{a}_3}{1+\tilde{\lambda}^2}\right)\langle\cdot \rangle_{\mathrm{bos.}} \ec
\end{gather}
where $\tilde{N}= \frac{2N\sin\pi\lambda}{\pi\lambda}$ and $\tilde{\lambda}=\tan\frac{\pi\lambda}{2}$. The normalized three-point functions
\begin{align}
\langle \cO_1\cO_2\cO_3\rangle^{\mathrm{norm.}}\equiv \frac{\langle \cO_1\cO_2\cO_3\rangle}{\sqrt{\langle \cO_1\cO_1\rangle \langle \cO_2\cO_2\rangle \langle \cO_3\cO_3\rangle}}\ec
\end{align}
are given by
\begin{align}
\langle J_0 J_0 J_s \rangle^{\mathrm{norm.}} = \tilde{N}^{-\frac{1}{2}}\langle\cdot\rangle^{\mathrm{norm.}}_{\mathrm{bos.}} \ecq \langle J_0 J_0 J_0 \rangle^{\mathrm{norm.}} = \tilde{N}^{-\frac{1}{2}}\frac{1 +\tilde{\lambda}^2 + \tilde{a}_3}{(1+\tilde{\lambda}^2)^{3/2}} \langle\cdot\rangle_{\mathrm{bos.}}^{\mathrm{norm.}}\ed \label{3pntnorm}
\end{align}
If we set $\tilde{a}_3 = (1+\tilde{\lambda}^2)(\sqrt{1+\tilde{\lambda}^2}-1)$ in \eqref{3pntnorm}, then all the 3-point functions $\langle J_0 J_0 J_s\rangle^{\mathrm{norm.}}$ for $s\geq 0$ differ from the free theory result by the same overall constant $\tilde{N}^{-1/2}$. In that case, the four-point function in the regular-boson theory was shown in \cite{Turiaci:2018nua,Aharony:2018npf} to be proportional to the free theory answer, up to a possible addition of three independent truncated solutions to crossing\cite{Heemskerk:2009pn}. The relation between $\tilde{a}_3$ and $\lambda_6$ can be found from the known three-point function \eqref{3pnt}, and is given by
\begin{align}
\tilde{a}_3 = \frac{1}{4}\tilde{\lambda}^2(1+\tilde{\lambda}^2)\left(3-\frac{\lambda_6}{8\pi^2\lambda^2}\right) \ed
\end{align}
In particular, we have $\tilde{a}_3 = (1+\tilde{\lambda}^2)(\sqrt{1+\tilde{\lambda}^2}-1)$ precisely when $\lambda_6$ is given by \eqref{lam6free}. This proves that the truncated solutions do not contribute in the regular-boson theory, in agreement with the results of \cite{Turiaci:2018nua} for the critical-fermion CFT.

Finally, it is easy to see that the Legendre transform of \eqref{4pnt} turns $G^{(4)}_6$ into contact-terms. The remaining non-trivial piece is proportional to the four-point function of $\bar{\psi}\psi$ in the theory of a free fermion. Again, this is in complete agreement with the results of \cite{Turiaci:2018nua} in the regular-fermion CS vector models, as expected from 3d bosonization duality.

\section*{Acknowledgements}

I would like to thank Ofer Aharony and Guy Gur-Ari for useful discussions.
This work was supported in part by the I-CORE program of the Planning and Budgeting Committee and the Israel Science Foundation (grant number 1937/12), by an Israel Science Foundation center for excellence grant, and by the Minerva foundation with funding from the Federal German Ministry for Education and Research.  

\appendix

\section{Proof of the Recurrence Relation}
\label{proof}

Let us now show that the recurrence relation \eqref{ansatz} solves the differential DS equation \eqref{diffDS}. Taking a derivative of \eqref{ansatz} gives\footnote{In this section, the $k_s$--dependence is left implicit ($V_{|p|}(p;k_s)\to V_{|p|}(p)$), to avoid clutter.}
\begin{align}
\frac{d}{dk_s}V_n(\{q_1,\ldots,q_n\}) &= a_n(\{q_1,\ldots,q_n\}) \frac{d}{dk_s}\left(V_1(\{q_1\})\cdots V_1(\{q_n\})\right) + \notag\\
&~ \sum_{p\in \cP'(n)} \left[(-1)^{|p|+1}\left( \prod_{i=1}^{|p|}b_{|p_i|}(p_i)\right)\frac{d}{dk_s}V_{|p|}(\bar{p})\right] \ec\label{dVn}
\end{align}
The identity we want to prove is obtained by plugging the differential DS equation \eqref{diffDS} on both sides of \eqref{dVn}:
\begin{align}
-4i\lambda\left(q_1+\cdots+q_n\right)\sum_{p\in P(n)} \cI(p)\prod_{i=1}^{|p|}V_{|p_i|}(p_i) = a_n(\{q_1,\ldots,q_n\})\frac{d}{dk_s}\left(V_1(\{q_1\})\cdots V_1(\{q_n\})\right)\notag\\
-4i\lambda\left(q_1+\cdots+q_n\right)\sum_{p\in \cP'(n)}(-1)^{|p|+1}\left(\prod_{i=1}^{|p|}b_{|p_i|}(p_i)\right)\sum_{p'\in \cP(\bar{p})}\cI(p')\prod_{i=1}^{|p'|}V_{|p'_i|}(p'_i) \ec\label{toprove}
\end{align}
where $\cI(p)$ is a function of $k_s$ and the external momenta defined in \eqref{int}.

As written, the equality \eqref{toprove} appears to be highly non-trivial. Our strategy will be to plug the recurrence relation on the LHS of \eqref{toprove} in a particular way, which makes this equality manifest. First, we decompose each monomial $\prod_i V_{|p_i|}$ on the LHS of \eqref{toprove} according to the rule\footnote{The identity \eqref{lemma} is obtained by expanding $y_1\cdots y_k = (x_1 - (x_1-y_1))(x_2-(x_2-y_2))\cdots(x_k-(x_k-y_k))$.}
\begin{align}
x_1 x_2 \cdots x_k &= \sum_{i=1}^k x_1 \cdots (x_i-y_i)\cdots x_k  \notag\\
&- \sum_{i_i<i_2}^{k} x_1\cdots (x_{i_1}-y_{i_1})\cdots(x_{i_2}-y_{i_2})\cdots x_k \notag\\
&+ \!\!\!\sum_{i_i<i_2<i_3}^{k}\!\!\! x_1\cdots (x_{i_1}-y_{i_1})\cdots(x_{i_2}-y_{i_2})\cdots (x_{i_3}-y_{i_3})\cdots x_k \notag\\
&\qquad\vdots\notag\\
& (-1)^{k+1} (x_1-y_1)(x_2-y_2)\cdots(x_k-y_k) \notag\\
&+ y_1y_2\cdots y_k \ec \label{lemma}
\end{align}
with $x_i\rightarrow V_{|p_i|}(p_i)$ and $y_i\rightarrow a_{|p_i|}(p_i)\left(V_1(\{p_{i1}\}) V_1(\{p_{i2}\})\cdots\right)$. After applying this decomposition, we plug in the recurrence relation \eqref{ansatz} once in all the $(x_i-y_i)$ terms. The terms $y_1\cdots y_k$ in \eqref{lemma} were already used in Section \ref{1PIvertices} to solve for the coefficients $a_n$, such that their sum matches $a_n(\{q_1,\ldots q_n\})\frac{d}{dk_s}\left(V_1(\{q_1\})\cdots V_1(\{q_n\})\right)$ on the RHS of \eqref{toprove}. Therefore, what is left to prove is
\begin{align}
\sum_{p\in\cP'(n)}\cI(p)\prod_{i=1}^{|p|}V_{|p_i|}(p_i)\bigg|_{\substack{\eqref{lemma}\text{ w.o. }y_1\cdots y_k\ec \\ (x_i-y_i)\to\eqref{ansatz}}} = \sum_{p\in\cP'(n)}(-1)^{|p|+1}\left(\prod_{i=1}^{|p|}b_{|p_i|}(p_i)\right)\sum_{p'\in P(\bar{p})}\cI(p')\prod_{i=1}^{|p'|}V_{|p'_i|}(p'_i)\ec \label{toprove2}
\end{align}
The notation on the LHS of \eqref{toprove2} means we decompose the $V_{|p_i|}$--monomials there according to \eqref{lemma} omitting the $y_1\cdots y_k$ terms, and plug in the recurrence relations in the ``$x_i-y_i$'' terms.\footnote{Note that the sum over $\cP(n)$ on the LHS of \eqref{toprove}, was replaced by one over $\cP'(n)$ on the LHS of \eqref{toprove2}. This is because the single partition $\{\{q_1\},\ldots,\{q_n\}\}$ of length $n$ decomposes as $x_1\cdots x_n=y_1\cdots y_n$ under \eqref{lemma}.} 

We claim that \eqref{toprove2} is manifestly satisfied. First, it is not difficult to see that the number of monomials in the vertex functions is the same on both sides. To see this we need
\begin{align}
\left|\cP(\text{set})\right| &= 2^{|\text{set}|-1} \ecq
N_{ni} = \binom{n-1}{i-1}\equiv \text{\# of } p\in \cP(n) \text{ of length }i\ec
\end{align}
from which it follows that 
\begin{align}
\sum_{p\in \cP(n)} x^{|p|} = \sum_{i=1}^{n} N_{ni} x^i = x(1+x)^{n-1}\ed
\end{align}
The number of monomials on the RHS of \eqref{toprove2} is therefore
\begin{gather}
\#\text{terms on RHS} = \sum_{p\in \cP'(n)}\left|\cP(\bar{p})\right|=\sum_{p\in \cP'(n)}2^{|\bar{p}|-1}=
\sum_{p\in \cP(n)}2^{|p|-1} - \left|\cP(n)\right|=3^{n-1}-2^{n-1} \ec
\end{gather}
where we used the fact that $|\bar{p}|=|p|$. Now let us consider the LHS of \eqref{toprove2}. For each $p\in \cP(n)$ we decompose $\prod_{i=1}^{|p|}V_{|p_i|}$ according to \eqref{lemma}. In each term of this decomposition, each $V_{|p_i|}$ either remains unaltered, or is transformed into an ``$x_i-y_i$'' term for which we plug in the recurrence relation. The recurrence relation for $V_{|p_i|}$ replaces it with a combination of $\left|\cP'(p_i)\right|$ lower-point vertices. Overall, each  $V_{|p_i|}$ in the monomial $\prod_{i=1}^{|p|}V_{|p_i|}$ is therefore transformed into a sum of $\left|\cP'(p_i)\right|+1=\left|\cP(p_i)\right|$ 1PI vertices. It follows that for each $p$ in the sum on the LHS of \eqref{toprove2}, there are $\prod_{i=1}^{|p|}\left|\cP(p_i)\right|-1$ monomials, where $1$ was subtracted to not count the possibility that $\prod_{i=1}^{|p|}V_{|p_i|}$ is left entirely unaltered. We conclude that
\begin{align}
\#\text{terms on LHS} &= \sum_{p\in \cP'(n)}\left[\left(\prod_{i=1}^{|p|}\left|\cP(p_i)\right|\right)-1\right] = \sum_{p\in \cP'(n)}\left[2^{\sum_{i=1}^{|p|}(|p_i|-1)}-1\right]\notag\\
&=\sum_{p\in \cP'(n)}\left[2^{n-|p|}-1\right]=\sum_{p\in P(n)}\left[2^{n-|p|}-1\right]=2^n\sum_{p\in P(n)}\left(\frac{1}{2}\right)^{|p|}-\left|\cP(n)\right|\notag\\
&=3^{n-1}-2^{n-1}= \#\text{terms on RHS}\ed
\end{align}

What is left is to find a one-to-one map between the monomials on both sides of \eqref{toprove2}. Specifically, for each $p\in\cP'(n)$ there are $|\cP(\bar{p})|$ monomials on the RHS of \eqref{toprove2}, and we will construct a sequence of maps between them and monomials on the LHS,
\begin{align}
(\text{RHS monomials})_{p\in\cP'(n)} \rightarrow \cP(\bar{p}) \xrightarrow{g_p}\cP(p)\xrightarrow{f_p} \cP(p)\otimes\cP^{(n)}(p)\rightarrow \text{LHS monomials} \ed \label{maps}
\end{align}
The mappings in \eqref{maps} are defined as follows:
\begin{enumerate}
	\item The first map simply associates each monomial in the sum over $p'\in\cP(\bar{p})$ on the RHS of \eqref{toprove2} with $p'$.
	\item For $\tilde{p}\in\cP(p)$ we define $g_p^{-1}(\tilde{p}) = \{\overline{\tilde{p}_1}, \overline{\tilde{p}_2},\ldots\}$. E.g., if $p=\{\{q_1,q_2\},\{q_3,q_4\},\{q_5\}\}\in\cP(5)$, and $\tilde{p}=\{\{\{q_1,q_2\},\{q_3,q_4\}\},\{\{q_5\}\}\}\in\cP(p)$, then
	 $g_p^{-1}(\tilde{p}) = \{\{q_1+q_2,q_3+q_4\},\{q_5\}\}\in \cP(\bar{p})$, with $\bar{p}=\{q_1+q_2,q_3+q_4,q_5\}$. The map $g_p$ is clearly bijective.
	 \item For $\tilde{p}\in\cP(p)$ we define $f_p(\tilde{p})\equiv(\tilde{p},\tilde{p}^{(n)})$, where $\tilde{p}^{(n)}\equiv \{\cup_j\tilde{p}_{1j},\cup_j\tilde{p}_{2j},\ldots\}\in\cP^{(n)}(p)\subset \cP(n)$. In the example of the previous bullet-point one finds $\tilde{p}^{(n)} = \{\{q_1,q_2,q_3,q_4\},\{q_5\}\}$. The map $f_p$ is one-to-one.
\end{enumerate}

We now construct the last map on \eqref{maps}. Let $(\tilde{p},\tilde{p}^{(n)})\in \cP(p)\otimes\cP^{(n)}(p)\subset\cP(p)\otimes\cP(n)$. On the LHS of \eqref{toprove2}, $\tilde{p}^{(n)}$ corresponds to a set of monomials
\begin{align}
\prod_{i=1}^{\left|\tilde{p}^{(n)}\right|} V_{|\tilde{p}^{(n)}_i|}(\tilde{p}^{(n)}_i) \bigg|_{\substack{\eqref{lemma}\text{ w.o. }y_1\cdots y_k \\ (x_i-y_i)\to\eqref{ansatz}}} \ed \label{LHS}
\end{align}
After plugging the recurrence relation \eqref{ansatz} in the decomposition \eqref{lemma}, \eqref{LHS} turns into a sum of monomials. We will pick the following monomial from this sum. First, focus on the term in the decomposition \eqref{lemma}, in which all  $V_{|\tilde{p}^{(n)}_i|}(\tilde{p}^{(n)}_i)$ in \eqref{LHS} are exchanged with ``$x_i-y_i$'', except if  $|\tilde{p}^{(n)}_i| = |\tilde{p}_i|$. Second, after plugging the recurrence relation for $V_{|\tilde{p}^{(n)}_i|}(\tilde{p}^{(n)}_i)$, pick the monomial corresponding to the partition $\tilde{p}_i\in\cP'(\tilde{p}^{(n)}_i)$. Specifically, if $|\tilde{p}^{(n)}_i| > |\tilde{p}_i|$ we replace
\begin{align}
V_{|\tilde{p}^{(n)}_i|}(\tilde{p}^{(n)}_i) \rightarrow (-1)^{|\tilde{p}_i|+1} \left(\prod_{j=1}^{|\tilde{p}_i|} b_{|\tilde{p}_{ij}|}(\tilde{p}_{ij})\right) V_{|\tilde{p}_i|}\left( \,\overline{\tilde{p}_i} \,\right)\ec \label{decomp}
\end{align}
and if $|\tilde{p}^{(n)}_i| = |\tilde{p}_i|$, we leave $V_{|\tilde{p}^{(n)}_i|}(\tilde{p}^{(n)}_i)$ unaltered. In addition, according to \eqref{lemma} we have to multiply by $(-1)^{d+1}$ where $d$ is the number of replacements $x_i\to x_i-y_i$ we made. To write this nicely, suppose we do use the rule \eqref{decomp} also on the vertices for which $|\tilde{p}^{(n)}_i| = |\tilde{p}_i|$. This results in multiplying the corresponding vertex by a minus sign:
\begin{align}
|\tilde{p}^{(n)}_i| = |\tilde{p}_i| \Rightarrow (-1)^{|\tilde{p}_i|+1} \left(\prod_{j=1}^{|\tilde{p}_i|} b_{|\tilde{p}_{ij}|}(\tilde{p}_{ij})\right) V_{|\tilde{p}_i|}\left(\,\overline{\tilde{p}_i}\,\right) = -V_{|\tilde{p}^{(n)}_i|}\left(\tilde{p}^{(n)}_i\right)\ed\label{wrongsign}
\end{align}
Indeed, if $|\tilde{p}^{(n)}_i| = |\tilde{p}_i|$ then $|\tilde{p}_{ij}|=1~\forall j$, because by definition $|\tilde{p}^{(n)}_i|=\sum_{j=1}^{|\tilde{p}_i|}|\tilde{p}_{ij}|\ge |\tilde{p}_i|$. This also implies that $\overline{\tilde{p}_i}=\tilde{p}^{(n)}_i$. Equation \eqref{wrongsign} then follows, because $b_1=-1$. We conclude that to pick the monomial we want we can simply apply the replacement rule \eqref{decomp} on the monomial in \eqref{LHS}, and multiply the result by $(-1)^{|\tilde{p}^{(n)}|+1} = (-1)^{|\tilde{p}^{(n)}|-d}(-1)^{d+1}$; the $(-1)^{d+1}$ is the sign on the $d+1$-th line of the identity \eqref{lemma}, and the $(-1)^{|\tilde{p}^{(n)}|-d}$ is there to cancel the unwanted signs in \eqref{wrongsign}.

To summarize, for each $(\tilde{p},\tilde{p}^{(n)})$ we choose a unique term on the LHS of \eqref{toprove2}, given by 
\begin{align}
\prod_{i=1}^{\left|\tilde{p}^{(n)}\right|} V_{\left|\tilde{p}^{(n)}_i\right|}\left(\tilde{p}^{(n)}_i\right) \rightarrow (-1)^{|\tilde{p}^{(n)}|+1}\prod_{i=1}^{\left|\tilde{p}^{(n)}\right|} \left[(-1)^{|\tilde{p}_i|+1} \left(\prod_{j=1}^{|\tilde{p}_i|} b_{|\tilde{p}_{ij}|}(\tilde{p}_{ij})\right) V_{|\tilde{p}_i|}\left(\overline{\tilde{p}_i}\right)\right] \ed \label{decompRule}
\end{align}
We have thus constructed a one-to-one map \eqref{maps} between $V_k$--monomials on both sides of \eqref{toprove2}. Re-instating all the factors the map is given by
\begin{align}
(-1)^{|p|+1}\cI(p')\left(\prod_{i=1}^{|p|}b_{|p_i|}(p_i)\right)\prod_{j=1}^{|p'|}V_{|p_i'|}(p_i') \,
\rightarrow\, (-1)^{|\tilde{p}^{(n)}|+1}\cI(\tilde{p}^{(n)})\prod_{i=1}^{\left|\tilde{p}^{(n)}\right|} \left[(-1)^{|\tilde{p}_i|+1} \left(\prod_{j=1}^{|\tilde{p}_i|} b_{|\tilde{p}_{ij}|}(\tilde{p}_{ij})\right) V_{|\tilde{p}_i|}\left(\overline{\tilde{p}_i}\right)\right]\ed\label{idmap}
\end{align}
The LHS of \eqref{idmap} is in one-to-one correspondence with $p\in\cP(n)$ and $p'\in\cP(\bar{p})$, while the RHS corresponds to $\tilde{p}=g_p(p')$ and $\tilde{p}^{(n)} = \{\cup_j\tilde{p}_{1j}, \cup_j\tilde{p}_{2j},\ldots\}$ ($g_p$ was defined bellow \eqref{maps}). Following the definitions, it is straightforward to see that both sides of \eqref{idmap} are equal. The minus signs work out:
\begin{gather}
(-1)^{|\tilde{p}^{(n)}|+1}\prod_{i=1}^{\left|\tilde{p}^{(n)}\right|} (-1)^{|\tilde{p}_i|+1} = (-1)^{|\tilde{p}|+1}\prod_{i=1}^{\left|\tilde{p}\right|} (-1)^{|\tilde{p}_i|+1}= 
(-1)^{|\tilde{p}|+1}\,(-1)^{|p|+|\tilde{p}|}=(-1)^{|p|+1}\ec
\end{gather}
where in the first equality we used the fact that $|\tilde{p}^{(n)}|=|\tilde{p}|$, and in the second equality we used $\sum_{i=1}^{|\tilde{p}|} |\tilde{p}_i|=|p|\ec\,\forall\tilde{p}\in \cP(p)$. Both properties follow directly from the definitions. For example the latter follows from the fact that $\tilde{p}\in \cP(p)\Rightarrow p=\cup_i\tilde{p}_i = \cup_{i,j}\left\{\tilde{p}_{ij}\right\}$, which also implies that 
\begin{align}
\prod_{i=1}^{|\tilde{p}^{(n)}|}  \prod_{j=1}^{|\tilde{p}_i|} b_{|\tilde{p}_{ij}|}(\tilde{p}_{ij}) = \prod_{i=1}^{|p|}b_{|p_i|}(p_i)\ed
\end{align}
Finally, it is easy to see from the definitions that $\overline{\tilde{p}_i}=p'_i$, $|\tilde{p}_i|=|p'_i|$ and $|\tilde{p}^{(n)}|=|\tilde{p}|=|p'|$ so (see also \eqref{int})
\begin{align}
\cI(\tilde{p}^{(n)})\prod_{i=1}^{|\tilde{p}^{(n)}|}V_{|\tilde{p}_i|}(\overline{\tilde{p}_i}) = \cI(p')\prod_{i=1}^{|p'|}V_{|p'_i|}(p'_i) \ed
\end{align}
This completes the proof of \eqref{toprove2}.

\bibliographystyle{utphys} 
\bibliography{ScalarCorrelators}

\end{document}